\documentclass[aip,amsmath,amssymb,reprint]{revtex4-1}

\usepackage{graphicx}
\usepackage{dcolumn}
\usepackage{bm}
\usepackage{xcolor}
\usepackage[utf8]{inputenc}
\usepackage[T1]{fontenc}
\usepackage{mathptmx}
\usepackage{etoolbox}

\makeatletter
\def\@email#1#2{%
 \endgroup
 \patchcmd{\titleblock@produce}
  {\frontmatter@RRAPformat}
  {\frontmatter@RRAPformat{\produce@RRAP{*#1\href{mailto:#2}{#2}}}\frontmatter@RRAPformat}
  {}{}
}%
\makeatother

\begin{document}

\title{Heralded single-pixel imaging with high loss-resistance and noise-robustness}

\author{Junghyun Kim}
 \affiliation{Agency for Defense Development, Daejeon 34186, South Korea}
\author{Taek Jeong}
 \affiliation{Agency for Defense Development, Daejeon 34186, South Korea}
\author{Su-Yong Lee}
 \affiliation{Agency for Defense Development, Daejeon 34186, South Korea}
\author{Duk Y. Kim}
 \affiliation{Agency for Defense Development, Daejeon 34186, South Korea}
\author{Dongkyu Kim}
 \affiliation{Agency for Defense Development, Daejeon 34186, South Korea}
\author{Sangkyung Lee}
 \altaffiliation{Authors to whom correspondence should be addressed: [Sangkyung Lee, sklee82@add.re.kr; Yong Sup Ihn, yong0862@add.re.kr]}
 \affiliation{Agency for Defense Development, Daejeon 34186, South Korea}
\author{Yong Sup Ihn}
 \altaffiliation{Authors to whom correspondence should be addressed: [Sangkyung Lee, sklee82@add.re.kr; Yong Sup Ihn, yong0862@add.re.kr]}
 \affiliation{Agency for Defense Development, Daejeon 34186, South Korea}
\author{Zaeill Kim}
 \affiliation{Agency for Defense Development, Daejeon 34186, South Korea}
\author{Yonggi Jo}
 \affiliation{Agency for Defense Development, Daejeon 34186, South Korea}

\date{\today}

\begin{abstract}
Imaging with non-classically correlated photon-pairs takes advantages over classical limits in terms of sensitivity and signal-to-noise ratio. However, it is still a challenge to achieve a strong resilience to background noise and losses for practical applications. In this work, we present heralded single-pixel imaging that is remarkably robust against bright background noise and severe signal losses. Using a strong temporal correlation between a photon-pair and joint measurement-based imaging method, we achieve the suppression of noise up to 1000 times larger than the signal and also demonstrate the correlation-induced SNR enhancement factor of over 200 against 70 times larger noise and a 90\% signal loss compared to non-time-gated classical imaging. Our work enables correlated imaging with a highly scalable photon capacity.
\end{abstract}

\maketitle

Non-classical correlations of light have been employed to enhance the sensitivity and signal-to-noise ratio (SNR) beyond the classical limits in quantum information processing \cite{NatPho11Giovannetti,NP10Kacprowicz,NatPho13Aasi,NatPho10Brida,NP17Slussarenko}. 
Especially, quantum illumination (QI) \cite{Sci08Lloyd, PRL08Tan, PRA21Lee, PRR21Jo}, which discriminates a presence and absence of a low reflectivity target using entangled states in noisy environments, outperforms the classical target sensitivity while quantum correlation survives without entanglement.
Recently, non-classical correlation-based imagings with QI have been experimentally demonstrated. 
In 2019, the target imaging with temporal correlation was demonstrated by means of raster-scanning \cite{PRA19England}. 
More recently, a full-field QI AND-imaging has been introduced \cite{SciAdv20Gregory}. 
It showed that the spatial correlation allows signal photons to be distinguished from uncorrelated noise photons by AND-operation of electron multiplying charged-couple devices (EMCCD). 
This QI AND-imaging achieved the noise rejection up to 5.8 and image contrasts up to a factor of 11 compared to classical imaging, and also further improved the modeling of noisy environments with thermal light.
However, due to the low temporal resolution of EMCCD ($\sim\mu$s), the measurement should be operated in a very low-photon regime to prevent saturation, which limited the range of acceptable noise and resulted in several hours of image construction time. To overcome long construction time, recent studies have shown a significantly reduced imaging time by using machine learning techniques and single-photon avalanche diode (SPAD) camera with a picosecond resolution \cite{Optica21Li, PRA21Defienne, npj20Ndagano}.
However, although imaging methods using non-classical sources have demonstrated the promising potential of noise suppression, imaging in an extremely severe condition, where the noise intensity is more than 100 times the signal intensity, has not yet been reported so far. Furthermore, unlike raster-scanning and multi-pixel imaging methods, single-pixel imaging with photon-pair illumination in various noise and loss conditions has not been investigated.

In this work, we perform the heralded single-pixel imaging with remarkable robustness in the presence of noise and losses. 
The proposed imaging scheme utilizes temporal correlations between photon-pairs downconverted from a continuous-wave (CW) pump laser. 
The use of single photon counting modules (SPCM) with sub-ns time resolution allows our imaging to achieve a scalable photon capacity without a saturation problem.
We use non-time-gated classical single-pixel imaging as a reference to validate our method in different noise and loss conditions.

Unlike imaging methods exploiting spatially correlated photon-pairs \cite{Optica21Li, PRA21Defienne, npj20Ndagano,PRA95Oittman,PRL05Valencia,SA19Defienne,SciAdv20Gregory,PRA19England}, single-pixel imaging (SPI), or computational ghost imaging (CGI), acquires spatial information of a target by illuminating it with time-varying patterns modulated by a spatial light modulator (SLM) \cite{OE20Gibson, PRA08Shapiro,SA17Liu,NP19Edgar,SR19Sun,APL20Yang,OE20Liu, SciRep17Sun}.
We define the $k$-th modulation pattern $P^{(k)}(i,j)$, which is represented by a 2D matrix with total $M$ entries, and the single-pixel detected intensity $I_{C(Q)}^{(k)}$, where the subscript $C$ ($Q$) denotes the classical (heralded) scheme. The image can be retrieved with the second-order correlation function between the patterns and intensities (see supplementary material):
\begin{equation}\label{Cov}
G^{(2)}(i,j)\: = \:\langle P^{(k)}(i,j)\: I_{C(Q)}^{(k)}\rangle - \langle P^{(k)}(i,j)\rangle \langle I_{C(Q)}^{(k)}\rangle
\end{equation}
where $i$ and $j$ are 2D pixel positions and $\langle...\rangle$ refers to the average value for the total pattern number $N$.
In the non-time-gated classical SPI scheme that signal photons illuminate the target in thermal background, the single-pixel detected photons can be given by $I_{C}^{(k)}=\eta_{o}'\eta_{s}\frac{1}{M}\sum_{i,j}P^{(k)}(i,j)\chi(i,j)n_{s}$ $+\eta_{s}n_{b}=\eta_{s}(\eta_{o}'\tilde{\chi}^{(k)}n_{s}+n_{b})$, where $n_{s}$ is the signal photon number injected into the SLM and $n_{b}$ is the number of noise photon. $\tilde{\chi}^{(k)}$ represents the overlapped portion between a pattern $P^{(k)}(i,j)$ and the target profile $\chi(i,j)$, $\eta_{o}'$ is an overall channel efficiency including losses, and $\eta_{s}$ refers to photon detection efficiency of an SPCM.

\begin{figure*}
\centering
\includegraphics[width = 0.8\textwidth]{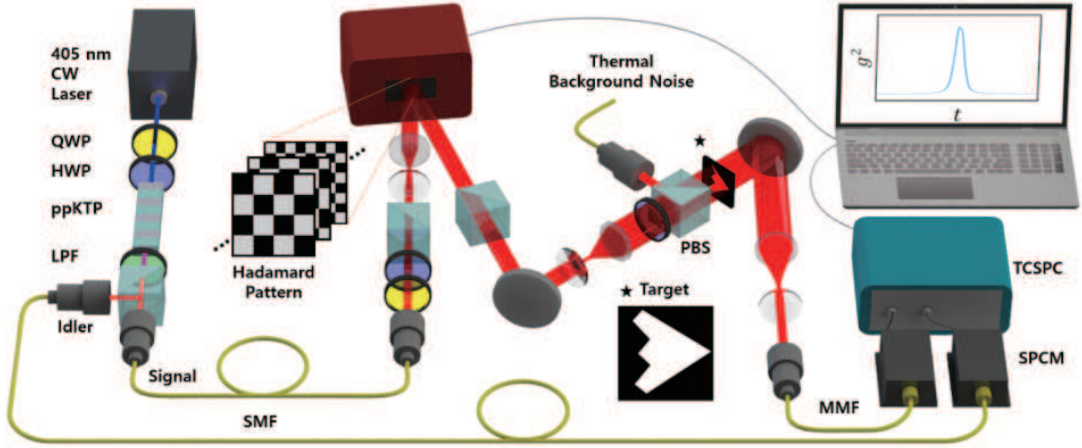}
\caption{Schematic of experimental setup. A 405 nm CW laser pumps a ppKTP crystal for type-II SPDC. While idler photons are directly measured for heralding, signal photons are projected onto 32 $\times$ 32 pixels Hadamard patterns by an SLM. In order to demonstrate the loss resistance and noise robustness, HWP and PBS are adjusted to control signal photon losses, and thermal noise photons are combined with signal photons at the PBS. The target object is a stealth-shaped aperture. After illuminating the target, both signal and noise photons are coupled to a multi-mode fiber (MMF) and analyzed with TCSPC. The target image is reconstructed by the correlation between the coincidence (or single) counts and modulation patterns. LPF, long-pass filter; TCSPC, time-correlated single photon counting.}\label{imagingsetup}
\end{figure*}

Here, when we take into account the noise photon number fluctuation effect to the classical SPI, the single-pixel detected photon count can be rewritten with its mean value and uncertainty:
\begin{equation}
I^{(k)} = \bar{I}^{(k)} + \delta I^{(k)}
\end{equation}
Then, the second-order correlation can be given by 
\begin{equation}\label{Cov2}
\begin{split}
G^{(2)}(i,j)=&\langle P^{(k)}(i,j) \bar{I}^{(k)}\rangle - \langle P^{(k)}(i,j)\rangle \langle \bar{I}^{(k)}\rangle\\
&+\langle P^{(k)}(i,j) \delta I^{(k)}\rangle - \langle P^{(k)}(i,j)\rangle \langle \delta I^{(k)}\rangle\\
=& \bar{G}^{(2)}(i,j) + \delta G^{(2)}(i,j)
\end{split}
\end{equation}
Here, $\bar{G}^{(2)}$ corresponds to the target image and $\delta{G}^{(2)}(i,j)$ shows the effect of noise photon fluctuation.
When the noise photon fluctuation is relatively small, the second term $\delta{G}^{(2)}(i,j)$ can be negligible and $\bar{G}^{(2)}(i,j)$ becomes dominant. 
For $\bar{I}^{(k)}=\eta_{s}(\eta_{o}'\tilde{\chi}^{(k)}\bar{n}_{s}+\bar{n}_{b})$, the contribution of mean photon number of noise, $\bar{n}_{b}$, is completely suppressed and we can obtain (see supplementary material)
\begin{equation}
\begin{split}
\bar{G}^{(2)}(i,j)&= \eta_o'\eta_s\bar{n}_s/4M \:\:\:\: \text{if} \:\:\:\: \chi(i,j) = 1 \\
& = 0 \:\:\:\:\:\:\:\:\:\:\:\:\:\:\:\:\:\:\:\:\:\:\:\:\:\:\:\:\:\:\:\:\:\: \text{if}\:\:\:\: \chi(i,j) = 0
\end{split}
\end{equation}
This implies thermal background noise can be quite rejected even in the non-time-gated classical SPI scheme where the photon number fluctuation is relatively small. 
However, as the noise level increases, it is not subtracted effectively and remains in $G^{(2)}(i,j)$ due to large photon number fluctuations.
If the noise level becomes significantly large, $\delta{G}^{(2)}$ overwhelms $\bar{G}^{(2)}$ so that the image suffers a considerable degradation.

In the heralded SPI scheme using an SPDC source, the temporal correlation of photon-pairs allows us to suppress the thermal background noise much further. 
When the SPDC source generates signal photons $n_{s}$ and idler photons $n_{i} (=n_{s})$ with a heralding efficiency $\eta_{h}$, the total photon number contributing to the joint detection with a coincidence window $T_{c}$ during a given acquisition time $\tau$, is written as $I_{Q}^{(k)}=\eta_{s}\eta_{i}(\eta_{h}\eta_{e}\eta_{o}\tilde{\chi}^{(k)}n_{s}+n_{b}n_{i}T_{c}/\tau)$, where $\eta_{i}$ is the detection efficiency of SPCM for idler photons, $\eta_{e}$ is a controllable target transmittance, and $\eta_{o}$ is the channel efficiency caused by experimental components ($\eta_o' = \eta_o\eta_e$). Therefore, ``signal loss'' in the rest of the paper corresponds to $\eta_e$. Here, the heralding efficiency is the ratio of coincidence counts rate to single counts rate of the SPDC source.

The first term of $I_{Q}^{(k)}$ represents the coincidence counts contributed by signal and idler photons, and the second term refers to the accidental counts between background noise and idler photons. 
It shows that the background noise can be significantly suppressed due to the small coincidence window $T_{c}$. 
Finally, heralded single-pixel images are obtained from $G^{(2)}(i,j)$ with coincidence counts $I_{Q}^{(k)}$ and modulation patterns $P^{(k)}(i,j)$.

The experimental schematic of the heralded SPI is shown in Fig. \ref{imagingsetup}.
Photon-pairs with the center wavelength of 810 nm are generated via type-II SPDC process by using a 10 mm-long periodically poled potassium titanyl phosphate (ppKTP) crystal with 10 $\mu$m polling period, pumped by a CW 405 nm laser. 
For SPDC source, the single count rates of signal and idler photons are almost same as 50 kcps/mW and the coincidence count rate is about 8 kcps/mW.

\begin{figure*}
\centering
\includegraphics[width = 0.8\textwidth]{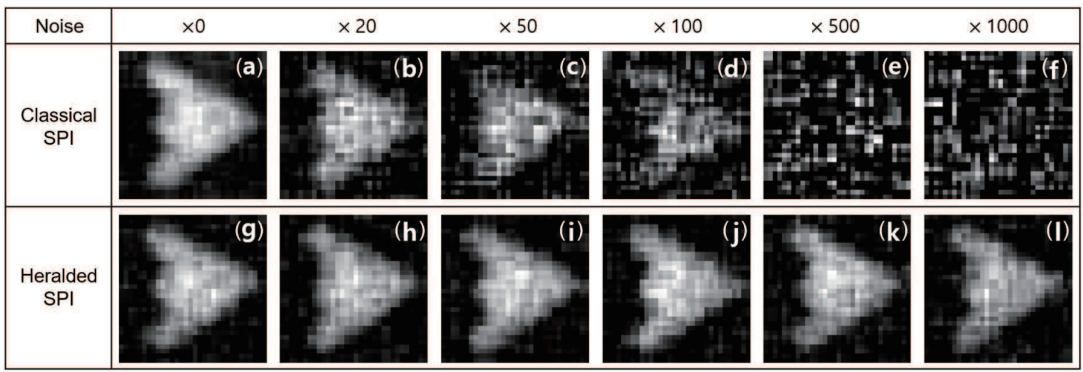}
\caption{Reconstructed classical (a-f) and haralded (g-l) SPI images in the presence of thermal background noise. As the noise level increases, classical SPI images are faded out, whereas heralded SPI images are insensitive to thermal background noise and clearly shown even when the target is subject to the background noise 1000 times stronger than the signal.}
\label{result1}
\end{figure*}

At 7.8 mW pump power, signal and idler photon counts are about 390 kcps.
The channel efficiency for signal photons is $\eta_{o}=5\%$, detector efficiencies for signal and idler are $\eta_s = \eta_i = 60\%$, and the heralding efficiency $\eta_h$ is 14\%.
Signal and idler photons are separated by PBS and coupled to single-mode fibers (SMF).
The idler photons are directly sent to an SPCM to herald signal photons. 
The signal beam first impinges on the liquid-crystal panel of the SLM (Thorlabs, EXULUS-HD1/M), which modulates the amplitude and phase of light fields with time-varying patterns.
We utilize the Hadamard patterns as a modulation basis, and images are constructed in 32 $\times$ 32 pixels resolution.
The mathematical definition of the Hadamard pattern is a square matrix obtained by reshaping a column of the Hadamard matrix \cite{SciRep17Sun}, which includes negative elements.
Theoretically, 1024 Hadamard patterns are sufficient to completely construct a $32\times32$ pixels image.
In the experiment, however, SLM only displays +1 or 0 (reflect or not) and two complementary patterns are required to describe one Hadamard pattern \cite{OE20Gibson}. 
We first make a modulation basis set with patterns $P^{(k)} = (J+H^{(k)})/2$ for $k = 1,2,...,1024$, where $J$ is a matrix of ones and $H^{(k)}$ is the $k$-th Hadamard pattern, and add inversed-patterns in the set which satisfy $P^{(k+1024)} = J-P^{(k)}$ for $k = 1,2,...,1024$. 
For one Hadamard pattern, two measurements are taken: one for the original pattern and the other for its inversed-pattern.
Therefore, the total number of acquisitions should be 2048 without compression. 
We apply compressive imaging method by only taking 700 patterns (350 original and 350 inversed) out of 2048 patterns based on the pattern ordering proposed in Ref. \cite{SciRep17Sun}.
With this approach, we achieve 3 times faster image reconstruction.
After being reflected from the SLM, the signal beam goes through the target having a stealth-shaped aperture.
To implement the loss and noise effects on the imaging, the transmittance of the target object ($\eta_{e}$) is controlled by a PBS and HWP, which simulates the signal losses, and thermal background noise photons overlap with signal photons through the PBS. 
The thermal noise beam is generated by focusing an independent 810 nm laser onto a rotating ground disk \cite{PL66Arecchi,PRL17Ihn}.
After illuminating the target, signal photons, mixed with thermal noise photons, are collected and coupled to an MMF and detected with an SPCM. 

For our heralded SPI, we first measure the cross-correlation function, $g^{(2)}{(\tau)}$, between the signal and idler photons and extract the coincidence counts from the peak value of the histogram.
Therefore, the imaging scheme does not depend on the path length difference between the signal and idler.
This is realistic because in real-world scenario, the target position is unknown and it is challenging to match two path lengths.
To ensure that all signal-idler photon-pairs contribute to coincidence detections, we set the coincidence window $T_{c}=650$ ps larger than twice of the temporal resolution of the detector ($\sim$300 ps) \cite{Optica19Liu}. 
Finally, the target image can be retrieved by calculating the second-order correlation $G^{(2)}(i,j)$ between the coincidence counts $I_{Q}^{(k)}$ and modulation patterns $P^{(k)}(i,j)$ after $N(=2\times350)$ times measurements. 
The acquisition time $\tau$ for each pattern is 1.5 s and pattern-switching time is set to 1 s. 
For the classical SPI, we simply measure the single counts $I_{C}^{(k)}$ of the signal-path instead of the coincidence count under the same condition. This means that all received signal-path photon counts during 1.5 s are recorded, and the idler photon counts are discarded. 

\begin{figure}
\centering
\includegraphics[width = 0.45\textwidth]{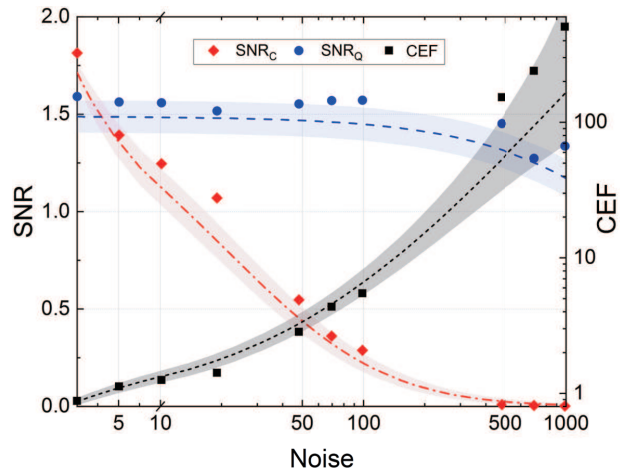}
\caption{SNRs of classical (red) and heralded (blue) images and CEFs (black) as a function of the noise level at no signal loss. The experimental data are obtained from the results of Fig. \ref{result1}. The theoretical values (lines and shaded regions) are calculated from Eq. \eqref{Cov} and \eqref{SNR} with $I^{(k)}\sim \text{Poi}$$(\lambda)$. The x-axis is linearly scaled from 0 to 10 and logarithmic-scaled from 10 to 1000.}
\label{SNR1}
\end{figure}

\begin{figure}
\centering
\includegraphics[width = 0.3\textwidth]{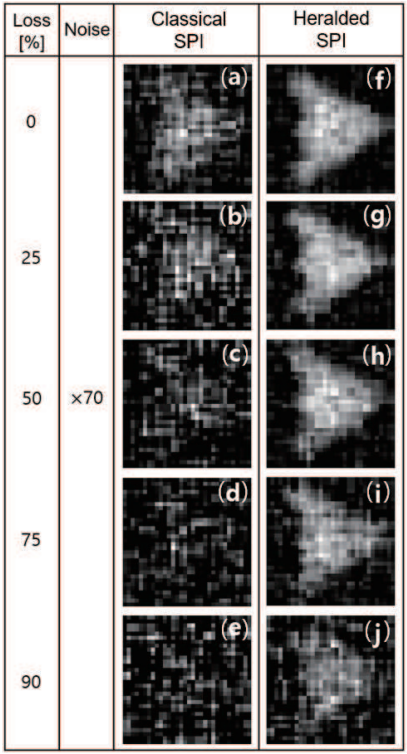}
\caption{Reconstructed (a-e) classical and (f-j) heralded SPI images in the presence of both background noise and losses. The signal photon loss at the target is varied from 0 to 90\% and the background noise level is fixed at 70 times larger than the averaged signal photon count rate of 7,800 cps.}
\label{result2}
\end{figure}

In Fig. \ref{result1}, we first compare the noise robustness of the classical and heralded SPIs under no signal losses ($\eta_{e}=1$) and illuminate the target with thermal background noise which is sufficient to increase 1000 times more than the average counts of received signal photons.
The averaged signal photon count rate is 7,800 cps. 
In classical images, the background noise having relatively small fluctuations are suppressed to some extent, as shown in Fig. \ref{result1} (a-c).
However, over the noise level of 50 times larger than the signal photon level, classical SPI images in Fig. \ref{result1} (d-f) suffer considerable degradations due to the large photon fluctuations. 
On the other hand, heralded SPI shows a remarkable noise-robustness and all images are clearly observed even at the 1000 times larger noise level, which demonstrates the highly scalable photon capacity. 
The strong temporal correlation of photon-pairs and small coincidence window effectively prevent noise contributions to the photon counting measurement. 
The corresponding SNRs of classical and heralded images are evaluated by the following equation \cite{SciAdv20Barzanjeh}:
\begin{equation}\label{SNR}
\text{SNR} = \frac{|\:\mu_{\text{T}} - \mu_{\text{B}}\:|^2}{2\:(\sigma_{\text{T}} + \sigma_{\text{B}})^2}
\end{equation}
where $\mu_{i}$ and $\sigma_{i}$ ($i = \text{T, B}$) denote the mean and standard deviation of gray-scale pixel values comprising of the target region and the background region of the image, respectively (see supplementary material).
To assess the robustness of heralded SPI compared to non-gated classical SPI, we define correlation-induced enhancement factor (CEF) as a ratio between the SNRs of two imaging schemes: CEF$=\frac{\text{SNR}_\text{Q}}{\text{SNR}_\text{C}}$.
In Fig. \ref{SNR1}, blue solid-circles and red diamonds present experimental data for SNR$_{\text{Q}}$ and SNR$_{\text{C}}$, respectively.
At the region below the noise level of 4.6, SNR$_{\text{C}}$ is slightly larger than SNR$_{\text{Q}}$ because the heralding efficiency is not unity in the heralded SPI, and the number of signal photons contributing to the image construction in the classical SPI is larger than that of the heralded SPI.
However, as the noise level increases up to 1000, SNR$_{\text{C}}$ diminishes quickly whereas SNR$_{\text{Q}}$ drops at a much slower rate and only decreases by 19\% of the noise-free SNR$_{\text{Q}}$ value.
At the noise level of 1000, we achieve the CEF of 500. 

\begin{figure}
\centering
\includegraphics[width = 0.45\textwidth]{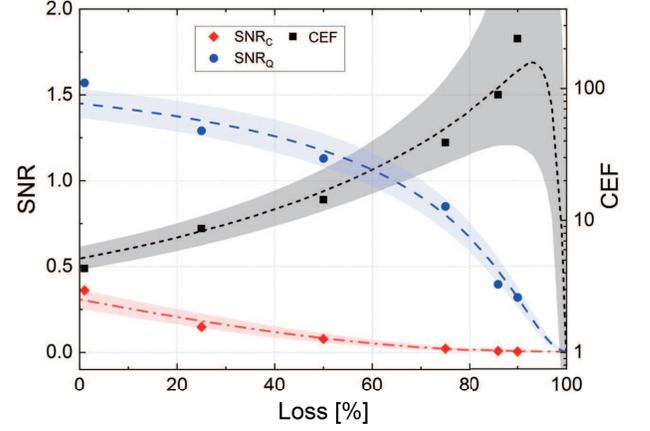}
\caption{SNRs of classical (red) and heralded (blue) images and CEFs (black) subject to the 70 times background noise level as a function of the losses at the target. The experimental data are obtained from the results of Fig. \ref{result2}. The theoretical values (lines and shaded regions) are simulated from Eq. \eqref{Cov} and \eqref{SNR} with $I_{C(Q)}^{(k)}$.}
\label{SNR2}
\end{figure}

The theoretical prediction of SNRs is calculated as follows.
We assume that the detected photon counts follow a Poisson distribution which is conditioned by the statistics of incoming photons due to the Poissonian photoelectric process in SPCM \cite{Optica19Liu,QOptics}. 
Here, we model the detection of photon counts with two steps: incoming photons are integrated during acquisition time, and then, the photoelectric process is conditioned by this integrated photon number. 
The quantum state of SPDC photon-pairs is two-mode squeezed vacuum state (TMSV) and if we trace out one mode, a single arm becomes a thermal state. 
Both signal and noise photons are thermal states and governed by Bose-Einstein distribution with different statistical mean $\bar{n}_s$ and $\bar{n}_b$.
During an acquisition time ($\tau=1.5$ s) much longer than the coherence times of signal photon ($\tau_s$) and noise photon ($\tau_b$), we obtain an integrated number of incoming photons. 
The Bose-Einstein statistics are valid within coherence time, and the integrated photon number can be considered as a sum of independent and identically distributed (i.i.d) random variables. 
The number of photons contributing to the photoelectric process is given as follows.
\begin{equation}\label{singleevents}
\begin{aligned}
\mu_C &=\:(n_{s,1} + n_{s,2} + ... + n_{s,{L_s}}) \\
&\:\:\:\:\:\:\:+ (n_{b,1} + n_{b,2} + ... + n_{b,{L_b}})\\
& = n'_s + n'_b
\end{aligned}
\end{equation}
where $L_w = \tau/\tau_w$ denotes the bin number for signal ($w=s$) and noise ($w=b$) photons, and $n_{s,j}$ and $n_{b,j}$ are random variables that follow Bose-Einstein distribution with mean $\bar{N}_s$=$\eta_{s}\eta_{e}\eta_{o}\tilde{\chi}^{(k)}\bar{n}_s$ and $\bar{N}_b$=$\eta_{s}\bar{n}_b$, respectively. 
According to the central limit theorem, each summation can be approximated to normal random variable $n'_w$ with mean $L_w\bar{N}_w$ and variance $L_w(\bar{N}_w+\bar{N}_w^2)$. 
Finally, the actual photon counts $I^{(k)}$ after the photoelectric process can be considered as a Poisson random variable whose mean value is $\mu_C$, i.e, $I^{(k)} \sim \text{Poi}(\lambda=\mu_C)$.
In the same manner, we can treat the single-pixel detected coincidence counts $I_Q^{(k)}$ as a Poisson random variable conditioned by a mean value $\mu_Q$. 
The idler photon number $n'_i$ follows a normal distribution with mean $L_s\eta_{i}\bar{n}_s$ and variance $L_s(\eta_{i}\bar{n}_s + (\eta_{i}\bar{n}_s)^2)$. 
Therefore, \begin{equation}\label{coincevents}
\mu_Q = \eta_{h}n'_s + n'_{i}n'_{b}T_{c}/\tau
\end{equation}
for heralded SPI.
In Fig. \ref{SNR1}, the red dot-dashed line and shaded region represent theoretical medians and uncertainties of SNR$_{\text{C}}$, respectively, which are obtained by substituting $I_{C}^{(k)} \sim \text{Poi}$$(\lambda=\mu_C)$ into Eq. \eqref{Cov} and calculating Eq. \eqref{SNR}.
For SNR$_{\text{Q}}$, theoretical values (blue dashed line and shaded region) are calculated from Eq. \eqref{Cov} and \eqref{SNR} with $I_{Q}^{(k)} \sim \text{Poi}$$(\lambda=\mu_Q)$.
From the ratio of two SNRs, theoretical values of CEF (black dot line and shaded region) are evaluated.
As a result, all experimental data are in good agreement with the theoretical predictions.  

In addition to noise effects, images can also be degraded by the signal photon loss due to low reflectivity of target that reduces the contrast between received photon numbers between patterns.
In Fig. \ref{result2}, we investigate the advantage of the heralded SPI, which is maintained when both noise and signal losses exist.
We control the signal photon loss from 0 to 90\% by changing the target transmittance from $\eta_{e}=$100 to 10\%.
Here, the background noise level is fixed to be 70 times larger than the signal level in the absence of photon losses.
It shows that the heralded SPI always outperforms the classical SPI in all transmittance range.
As the signal photon loss increases up to 90\%, classical images become totally broken and indistinguishable from the image background.
On the other hand, heralded images are highly robust to the noise and losses, and provide a perceptible target image even at the 70 times stronger noise and 90\% signal loss.
As a result, we achieve the CEF up to a factor of 240 compared to classical imaging as shown in Fig. \ref{SNR2}.  

In conclusion, we have demonstrated the heralded SPI that is highly robust to a noisy and lossy environment.
We showed that the strong temporal correlation and joint measurement between downconverted photon-pairs could provide a substantial improvement and scalability to correlated imaging schemes.
This cross-correlation measurement scheme does not rely on the optical path length difference between the pair.
Finally, we note that our scheme can be further improved by the use of a highly reflective digital micro-mirror device (DMD) with kHz frame rates.
Our work provides the way towards practical applications for remote quantum sensing and imaging with QI.

\section*{Supplementary material}
See supplementary material for Eq.(1), Eq.(4), and SNR calculation.

\begin{acknowledgments}
This work was supported by a grant to Defense-Specialized Project funded by Defense Acquisition Program Administration and Agency for Defense Development.
\end{acknowledgments}

\section*{Conflict of interest}
The authors have no conflicts to disclose.

\section*{Data Availability}
The data that support the findings of this work are available from the corresponding authors upon reasonable request.

\section*{References}

\end{document}